\documentclass[prl,twocolumn,showpacs,twoside,showpacs]{revtex4}

\usepackage{graphicx}
\usepackage{dcolumn}
\usepackage{bm}

\begin{document}

\bibliographystyle{prsty}
\title{On the isoscalar-isovector splitting of pygmy dipole structures}
\author{N. Paar$^1$}
\email{npaar@phy.hr}
\author{Y. F. Niu$^{1,2}$}
\author{D. Vretenar$^1$}
\author{J. Meng$^{2,3}$}
\affiliation{$^1$Physics Department, Faculty of Science, University of Zagreb, 
Croatia}
\affiliation{$^2$State Key Laboratory for Nuclear Physics and Technology, School of Physics, Peking University, Beijing 100871, China}
\affiliation{$^3$School of Physics and Nuclear Energy, Beihang University, Beijing 100083, China}

\date{\today}
\begin{abstract}
The electric dipole response of $^{140}$Ce is investigated using the 
fully consistent relativistic quasiparticle random phase approximation. By analyzing the 
isospin structure of the E1 response, it is shown that the 
low-energy (pygmy) strength separates into two segments with different 
isospin character. The more pronounced pygmy structure at lower energy 
is composed of predominantly isoscalar states with surface-peaked 
transition densities. At somewhat higher energy the calculated E1 strength 
is primarily of isovector character, 
as expected for the low-energy tail of the giant dipole resonance.
The results are in qualitative agreement with those obtained in 
recent  $(\gamma,\gamma')$ and $(\alpha,\alpha'\gamma)$
experiments, and provide a simple explanation for the 
splitting of low-energy E1 strength into two groups of states
with different isospin structure and radial dependence of the 
corresponding transition densities.
\end{abstract}
\pacs{ 21.10.Gv,21.30.Fe,21.60.Jz,24.30.Cz}
\maketitle
\date{today}

Exotic modes of collective excitation represent unique
structure phenomena in nuclear many-body systems with 
a pronounced asymmetry in the number of protons and neutrons \cite{PVKC.07}.
A representative example is the  
pygmy dipole resonance (PDR), a low-energy mode 
that  corresponds to a resonant oscillation of the weakly-bound neutron 
skin against the isospin saturated proton-neutron core. 
Even though high-resolution data on low-energy dipole
states have been obtained in recent  $(\gamma,\gamma')$ experiments 
\cite{Sav.06,Sav.08,Vol.06,Sch.08,Oze.08,Oze.07,Zil.02},
the details of the underlying nature and dynamics of the PDR
remain largely unknown.  Recent experiments
with radioactive ion beams have extended the studies 
of low-lying dipole response towards exotic nuclear
systems with pronounced neutron excess~\cite{Adr.05,Wie.09}.
The occurrence of  PDR has also been investigated 
using a variety of theoretical approaches and models \cite{PVKC.07}.  
More recent studies have made use of the
quasiparticle RPA plus phonon coupling~\cite{Sar.04},
the quasiparticle-phonon model~\cite{Tso.08}, 
the relativistic RPA~\cite{Pie.06,Lia.07} and
QRPA~\cite{PNVR.05}, as well as the relativistic quasiparticle
time blocking approximation~\cite{Lit.07,Lit.08}.

A recent  $(\alpha,\alpha'\gamma)$ study of electric-dipole 
excitations below the particle threshold in the semi-magic nucleus 
$^{140}$Ce \cite{Sav.06}, has shown that 
the PDR structure separates into two parts. By comparing the 
observed low-lying strength to results from photon scattering 
experiments on the same target \cite{Zil.02}, it was noticed 
that the lower part of the PDR is excited both in $(\alpha,\alpha'\gamma)$
and $(\gamma,\gamma')$ scattering, whereas the higher-energy
component seems not to be excited by $\alpha$-particles.

In this letter we analyze the isospin properties and transition 
densities of low-lying dipole excitations in $^{140}$Ce. 
Similar aspects of the PDR will be considered as in Ref.~\cite{Sav.06} , 
but from a theoretical point of view. The goal is to explore in simple 
terms, i.e. by analyzing the isotopic and spatial dependence of transition densities 
for dipole excitations, the splitting of the PDR structure into two groups of states, 
excited either by both the nuclear part of the $\alpha$-nucleon interaction and 
photons, or only photons, respectively.

The present analysis uses the fully self-consistent relativistic
quasiparticle random phase approximation (RQRPA) based on the
Relativistic Hartree-Bogoliubov model (RHB) \cite{Paa.03}. 
Details of the formalism can be found in Refs.~\cite{Paa.03,PVKC.07}.  
In the RHB+RQRPA model the effective interactions are implemented in 
a fully consistent way. In the particle-hole channel effective Lagrangians
with density-dependent meson-nucleon couplings are employed~\cite{PNVR.04},
and pairing correlations are described by the pairing part of the finite-range 
Gogny interaction~\cite{BGG.91}. 
Both in the $ph$ and $pp$ channels, the same interactions are used in the RHB
equations that determine the canonical quasiparticle basis, and in the
matrix equations of the RQRPA. 

The full set of RQRPA equations is solved by diagonalization. The result are
excitation energies $E_{\lambda}$ and the corresponding
forward- and  backward-going amplitudes, $X^{\lambda}$ and $Y^{\lambda}$, respectively,
that are used to evaluate the transition strength: 
\begin{eqnarray}
B^T_J(E_{\lambda}) & = & \frac{1}{2J_{i}+1}
\bigg\vert \sum_{\mu\mu'} \bigg\{ X^{\lambda, J0}_{\mu\mu'} \langle
\mu || \hat{Q}^T_J || \mu' \rangle \nonumber \\
& + &~(-1)^{j_{\mu}-j_{\mu'}+J} \, Y^{\lambda, J0}_{\mu\mu'}
\, \langle \mu' || \hat{Q}^T_J || \mu \rangle \,
\bigg\} \nonumber \\
& &(u_{\mu}v_{\mu'}+~(-1)^{J}v_{\mu}u_{\mu'})
\bigg\vert ^2 \quad ,
\label{strength}
\end{eqnarray}
where $\mu$ and $\mu'$ denote canonical single-nucleon states,
and $u$, $v$ are the corresponding occupation factors.
Discrete spectra are averaged with a Lorentzian distribution
of arbitrary width ($\Gamma$=1.5 MeV in the present calculation), 
and the electric E1 response is calculated for the isovector dipole operator:
\begin{equation}
\hat{Q}_{1 \mu}^{T=1} \ = \frac{N}{N+Z}\sum^{Z}_{p=1} r_{p}Y_{1 \mu}
- \frac{Z}{N+Z}\sum^{N}_{n=1} r_{n}Y_{1 \mu} \; .
\end{equation}
Particularly
relevant for the present study are the neutron and proton radial transition densities
$\delta\rho_{n,p}$(r), which describe nucleon density fluctuations 
induced by an external field~\cite{Ser.83}.  For instance, by analyzing 
if $\delta\rho_{n}$(r) and $\delta\rho_{p}$(r) for a particular state 
appear to be in phase over some extended region within the nuclear volume,
states with predominant isoscalar structure can be identified in
the RQRPA dipole strength distribution (Eq.~(\ref{strength})). 
\begin{figure}
\centerline{
\includegraphics[scale=0.37,angle=0]{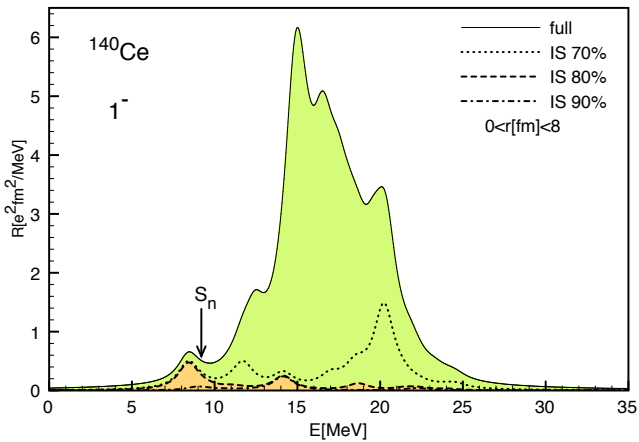}
}
\caption{The total RHB+RQRPA E1 transition strength for 
$^{140}$Ce (solid). The dotted, dashed, and dot-dashed curves connect states 
with predominant (at least  70\%, 80\% and 90\%, respectively) isoscalar (IS) components,  
identified by analyzing the corresponding proton and neutron transition densities 
over the radial interval $[0,8]$ fm. The vertical arrow denotes the one-neutron 
separation energy $S_n = 9.2$ MeV.}
\label{fig1}
\end{figure}

The solid curve in Fig.~1 represents the total RHB+RQRPA E1 transition strength for 
$^{140}$Ce, calculated with the DD-ME2~\cite{LNVR.05} parameterization of the effective 
Langrangian. In addition to the pronounced collective isovector giant dipole resonance 
(GDR) peaked at $\approx 15$ MeV, we notice an enhancement of E1 strength in the 
low-energy region below neutron threshold, indicated by the vertical arrow. This structure could 
be attributed to a PDR. The dotted, dashed, and dot-dashed curves illustrate the isospin 
structure of the E1 strength distribution. These curves connect states with predominant 
(at least  70\%, 80\% and 90\%, respectively) isoscalar components,  
identified by analyzing the corresponding proton and neutron transition densities 
over the radial interval $[0,8]$ fm. When for a particular state the proton and neutron 
transition densities are found to be in phase over more than 70\% of this range of the 
radial coordinate, the state is denoted IS 70\%, and analogously for the states 
IS 80\% and IS 90\%. For the low-energy excitations around the pronounced
peak at 8.4 MeV the proton and neutron transition densities are in phase 
over more than 80\% of the radial interval $[0,8]$ fm, i.e. these states are 
predominantly isoscalar.  Some isoscalar states are also found at 
excitation energy E$>$10 MeV but in that interval, obviously, 
isovector collective motion dominates. The analysis of the isospin structure of 
the E1 transition strength in Fig.~1 indicates  that most of the low-energy
strength is of isoscalar character, and therefore sensitive
to isoscalar probes such as $\alpha$-particles. States at higher energy, 
however, would probably not be excited in $(\alpha,\alpha'\gamma)$ experiments
because of their predominant isovector character. Photon scattering, on the 
other hand, should excite dipole states over a broader low-energy interval, 
but $(\gamma,\gamma')$ experiments have been restricted to the region 
below the neutron separation energy. Indeed, results of RQRPA 
calculations \cite{PNVR.05,PVKC.07} indicate that 
part of the PDR strength may be missing in $(\gamma,\gamma')$ experiments, 
i.e., these calculations predict PDR strength around and also
above the neutron threshold. Some progress in recovering part of
the unresolved E1 strength has already been reported~\cite{Oze.08},
but future experimental studies across the neutron threshold may 
be necessary to account for the missing low-lying E1 strength.

\begin{figure}
\centerline{
\includegraphics[scale=0.37,angle=0]{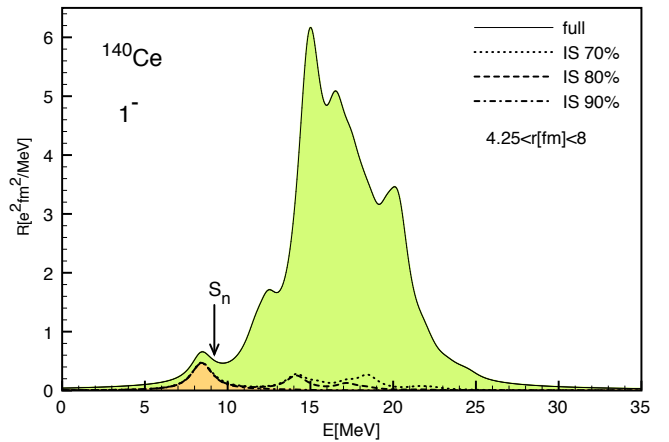}
}
\caption{Same as in Fig.~1, but the dotted, dashed, and dot-dashed 
curves are determined from the radial dependence of the neutron 
and proton transition densities in the surface region: 4.25 $<$ r $<$ 8 fm. }
\label{fig2}
\end{figure}

The photon field interacts with a nucleus as a whole, whereas $\alpha$-particles 
interact mainly with the nuclear surface, inducing isoscalar transitions with 
surface-peaked transition densities. It is, therefore, interesting to analyze in more  
detail the radial structure of the transition densities for the dipole states 
shown in Fig.~1. When the identification of states with predominant isoscalar 
components is based on the behavior of the 
corresponding neutron and proton transition densities in the radial interval 
$[0,4.25]$ fm, i.e., in the volume of the nucleus, one finds that these 
states are distributed over a wide energy region, but not below E $\approx 10$ MeV. 
Fig.~2, on the other hand, displays the same analysis of the 
E1 transition strength, but the predominantly isoscalar states are identified 
by considering the in-phase or out-of-phase behavior of the corresponding 
neutron and proton transition densities in the surface region 4.25 $<$ r $<$ 8 fm. 
In this case transitions with predominant isoscalar character are located mainly
in the low-energy region below 10 MeV. One can conclude that the low-energy
isoscalar structure peaked at 8.4 MeV corresponds mainly to surface excitations.
Low-lying isoscalar states with surface-peaked transition densities are precisely 
those expected to be induced in $\alpha$-scattering experiments.

\begin{figure}
\centerline{
\includegraphics[scale=0.35,angle=0]{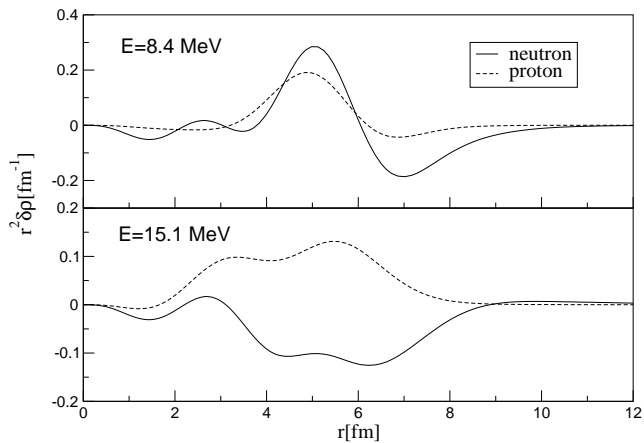}
}
\caption{The RQRPA neutron and proton transition densities for the peaks 
at 8.4 MeV and 15.1 MeV excitation energy in $^{140}$Ce.
}
\label{fig3}
\end{figure}

In Fig.~3 we plot the neutron and proton transition densities for the peaks 
at 8.4 MeV and 15.1 MeV. For the main peak at 15.1 MeV the transition 
densities display a radial dependence that is characteristic for the isovector 
GDR: the proton and neutron densities oscillate with opposite phases. 
The dynamics of the state at 8.4 MeV is different: the proton and neutron 
transition densities are in phase in the nuclear interior and 
there is almost no contribution from the protons
in the surface region. This low-energy state does not belong to
statistical E1 excitations sitting on the tail of the GDR.
Fig.~4 explores in more detail the structure of the transition matrix elements 
for the PDR state at 8.4 MeV. 
The matrix elements (i.e. each term in the sum over $\mu$ and $\mu'$ in Eq.~(1), 
representing the contribution of the corresponding $2qp$ configuration to the transition 
strength) are plotted as a function of the unperturbed energy of the
corresponding $2qp$ configurations. Proton and neutron 
matrix elements are displayed separately. First, we note that the state at 8.4 MeV 
is characterized by a coherent superposition (matrix elements of the same sign) 
of more than ten neutron $2qp$ configurations. The largest matrix elements 
correspond to the following neutron transitions (in order of increasing $2qp$ 
unperturbed energy): 
$3s_{1/2} \rightarrow 3p_{3/2}$, $2d_{3/2} \rightarrow 3p_{3/2}$,
$3s_{1/2} \rightarrow 3p_{1/2}$, $2d_{3/2} \rightarrow 3p_{1/2}$,
$1h_{11/2} \rightarrow 1i_{13/2}$, $2d_{5/2} \rightarrow 3p_{3/2}$,
$1g_{7/2} \rightarrow 1h_{9/2}$, $1h_{11/2} \rightarrow 2g_{9/2}$,
$1g_{7/2} \rightarrow 2f_{5/2}$.
Second, from the analysis of the  QRPA amplitudes 
$(X^\lambda_{2qp})^2$-$(Y^\lambda_{2qp})^2$ it follows 
that the total contribution of proton $2qp$ excitations to the state at 8.4 MeV 
is only 9\%, well below the ratio Z/N expected for a GDR state, whereas 
91\% of the total amplitude corresponds to neutron transitions.

\begin{figure}
\centerline{
\includegraphics[scale=0.37,angle=0]{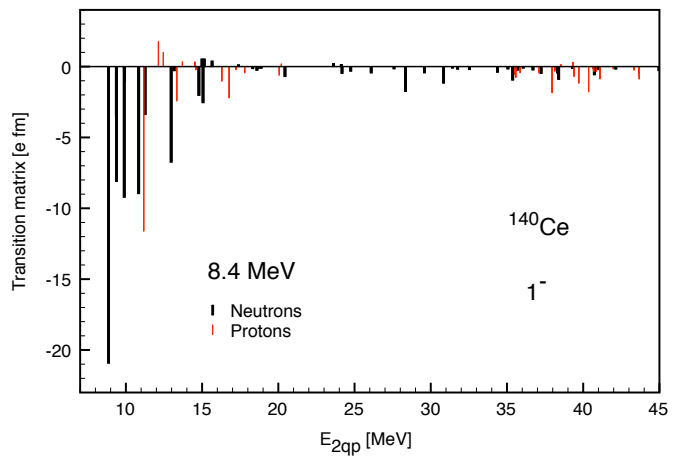}
}
\caption{The RQRPA neutron and proton transition matrix elements for the state at 
8.4 MeV in $^{140}$Ce, as a function of the unperturbed energy 
of the corresponding $2qp$ configurations.}
\label{fig4}
\end{figure}

The relativistic QRPA accurately reproduces experimental 
excitation energies of the GDR, whereas the calculated PDR are only in
qualitative agreement with available data. 
In the particular example of $^{140}$Ce, the predicted PDR 
structure is calculated more than 2 MeV above the pygmy structure seen in 
$(\gamma,\gamma')$ and $(\alpha,\alpha'\gamma)$ experiments \cite{Sav.06}. 
The principal reason for  this discrepancy lies in the fact that  
relativistic mean-field effective interactions, e.g. 
DD-ME2, have relatively low effective nucleon masses, typically $m^* < 0.7~m_N$, 
where $m_N$ denotes the free nucleon mass. A larger effective mass, 
i.e. a higher density of states around the Fermi surface, would lower the 
excitation energy of states with predominant isoscalar components, 
e.g. the PDR. Unfortunately, it is not simple to increase the effective nucleon mass  
in relativistic mean-field models, because this quantity is strongly constrained 
by the Dirac mass, which represents a measure of the strength of the 
spin-orbit single-nucleon potential \cite{VNR.02,NVR.05}. In addition, our study is 
limited to a simple QRPA and does not include effects of 
coupling to low-energy surface phonons. This coupling will also enhance 
the effective nucleon mass. 
The analysis of PDR can be extended by introducing the coupling of
two-quasiparticle excitations to collective vibrations, e.g.
using the quasiparticle phonon model~\cite{Tso.08}, or the
relativistic quasiparticle time blocking approximation~\cite{Lit.07}. 
In Ref.~\cite{Lit.07} it has been shown that the effect of two-phonon 
admixtures is a small shift of PDR states ($\leq 1$ MeV) to lower 
excitation energies but, although the PDR calculated in the extended space 
contains sizable two-phonon admixtures, it basically retains a one-phonon 
character and the pygmy dynamics is not modified by the coupling to 
low-energy surface vibrations.
\begin{figure}
\centerline{
\includegraphics[scale=0.37,angle=0]{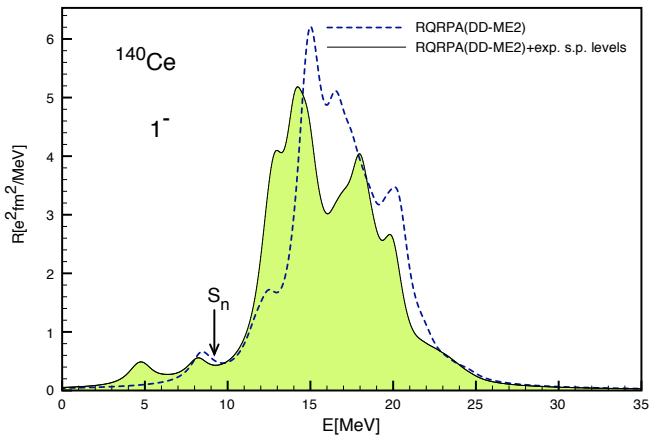}
}
\caption{The total RHB+RQRPA E1 transition strength for 
$^{140}$Ce in comparison with calculation using experimental
neutron single-particle energies.}
\label{fig5}
\end{figure}

To illustrate how sensitive PDR is to the effective nucleon mass, in Fig.~5 
we display the fully self-consistent RQRPA E1 strength calculated with the DD-ME2 
interaction, in comparison to the E1 strength obtained with the same effective 
interaction, but with DD-ME2 neutron single-particle energies 
below and above the $N=82$ shell gap replaced with experimental values. 
The empirical $N=82$ gap is much smaller than the one 
predicted by DD-ME2, i.e. the empirical effective neutron mass is larger.  
By using experimental neutron single-particle energies one takes into account, 
at least on the level of single-nucleon energies, the effect of coupling to 
collective vibrations but, of course, the calculation is no longer self-consistent. 
Nevertheless, we notice that the enhancement of the effective 
neutron nucleon mass has a pronounced effect  especially on the PDR structure. 
The E1 strength below 10 MeV is fragmented into two structures, with the lower 
one centered at $\approx 5$ MeV and higher one peaked at $\approx 8$ MeV,
in much better agreement with data~\cite{Sav.06}. 
An analysis of the isospin content of these structures, based on proton and neutron 
transition densities, shows that isoscalar components are more pronounced in the 
lower structure at  $\approx 5$ MeV. We have also verified that the same pattern is 
calculated for $^{132}$Sn, when DD-ME2 single nucleon energies are replaced 
by empirical values.

In conclusion, the fully consistent relativistic QRPA has been used to 
calculate the electric dipole response of $^{140}$Ce. An analysis of the 
isospin structure of E1 strength, based on the radial dependence of the 
microscopic proton and neutron transition densities, has shown that the 
low-energy strength separates into two segments with qualitatively different 
isospin character. The more pronounced PDR structure at lower energy 
is composed of predominantly isoscalar states with surface-peaked 
transition densities. At energies above the main PDR peak the E1 strength 
is primarily of isovector character, with non-surface peaked transition densities, 
as expected for the low-energy tail of the giant dipole resonance.
The results are in qualitative agreement with those obtained in 
recent  $(\gamma,\gamma')$ and $(\alpha,\alpha'\gamma)$
experiments, and provide a simple explanation for the 
splitting of low-energy E1 strength into two groups of states, 
excited either by both the nuclear part of the  $\alpha$-nucleon interaction and 
photons, or only photons, respectively.

This work was supported by the Unity through 
Knowledge Fund (UKF Grant No. 17/08) and MZOS - project 1191005-1010. Y. F. Niu 
acknowledges support from the National Foundation for Science, Higher Education and
Technological Development of the Republic of Croatia. The work of J.M and D.V. was 
supported in part by the Chinese-Croatian project ``Nuclear structure far from stability". 

%

\end{document}